# The differing meanings of indicators under different policy contexts. The case of internationalisation

## Nicolas Robinson-Garcia[1] and Ismael Ràfols[2,3]

[1] Delft Institute of Applied Mathematics (DIAM), TU Delft, Delft, Netherlands
[2] INGENIO (CSIC-UPV), Universitat Politècnica de València, Valencia, Spain
[3] Centre for Science and Technology Studies (CWTS), Leiden University, Leiden, Netherlands

## Abstract

In this chapter we build upon Moed's conceptual contributions on the importance of the policy context when using and interpreting scientometric indicators. We focus on the use of indicators in research evaluation regarding internationalisation policies. The globalization of higher education presents important challenges to institutions worldwide, which are confronted with tensions derived from the need to respond both, to their local necessities and demands while participating in global networks. In this context, indicators have served as measures for monitoring and enforcing internationalisation policies, in many cases interpreting them regardless of the policy context in which they are enforced. We will analyse three examples of indicators related to internationalisation. The first one is about international collaborations, under the assumption that a greater number of internationally co-authored publications will benefit a national science system as it will result in higher citation impact. The second one relates to the promotion of English language as the dominant language of science. The third case analyses how different policy contexts shape the selection and construction of indicators, sometimes in a partial way which does not properly reflect the phenomenon under study. The examples illustrate that the interpretation and policy implications of the 'same' S&T indicators differ depending on specific contexts.

## Introduction

The development and growth of the field of evaluative scientometrics cannot be understood without the fundamental contributions of Henk Moed. Along with his colleagues at the University of Leiden Centre for Science and Technology Studies (CWTS), he became a key player on establishing the basic pillars for the use of bibliometric indicators for research assessment (Moed, Bruin, & Leeuwen, 1995; Moed, Burger, Frankfort, & Van Raan, 1985). Moed's work has been characterized by a critical notion on the use of indicators. He was one of the first to point out potential problems derived from the use of the Impact Factor for research assessment (Moed & van Leeuwen, 1996; Moed & Van Leeuwen, 1995), or the limitations of scientometrics when assessing the citation impact of non-English literature (Leeuwen, Moed, Tijssen, Visser, & Raan, 2001) among others. His two single-authored books ( Moed, 2005, 2017b), essential readings for anyone interested on the field, are characterized by an open-minded and pedagogical tone which reflects a critical and constructive view of evaluative scientometrics.

In his latest book, Moed proposes shifting away from a 'narrow' evaluative use of indicators to a more analytical one. He warns that "[t]o the extent that in a practical application an evaluative framework is absent or implicit, there is a vacuum, that may be easily filled either with ad-hoc arguments of evaluators and policy makers, or with un-reflected assumptions underlying informetric tools" (Moed, 2017, p. 29). In his view, the selection of indicators should be made within the 'policy context' in which they are going to be implemented (Moed & Halevi, 2015). Building upon this body of work, in this chapter we aim at further exploring this 'analytical' perspective on the use of scientometrics. We stress that context will not only provide the appropriate framework for the selection of indicators, but also for their interpretation, moving from a universal interpretation of indicators to a context-dependent one.

At this point, it is important to distinguish between policy context and adapting the indicators to a given context, what Waltman (2019a, 2019b) refers to as 'contextualised scientometrics'. In the latter case, context is understood as a means to ensure transparency, and facilitate a better understanding on how the indicator is constructed and adapted to specific fields, countries or languages. The purpose in 'contextualised scientometrics' is to allow the user to grasp the limitations and biases inherent to scientometric indicators so that they are not misinterpreted due to technical and conceptual assumption on what the indicator is measuring. This is the line of thought followed by Gingras (2014) when defining the three desirable characteristics of a well-designed indicator: 1) adequacy for the object it measures, 2) sensitivity to the intrinsic inertia of the object, and 3) homogeneity of the dimensions of the indicator. However, the focus here is in policy context, which has to do with the understanding of the purpose of the assessment, the selection of the indicator and its interpretation based on broader social or policy factors which may be crucial to understand what the indicator is actually portraying.

To illustrate the importance of policy context when interpreting scientometric indicators, in this chapter we will focus on their use for studying the effects of internationalisation policies in science. The aim is to highlight how a de-contextualised use of scientometric indicators can work against the expected goals for which indicators were originally introduced. The phenomenon of globalization in science gives us a good example to explore such ambiguity, as many countries have turned their attention towards scientometrics in order to implement internationalisation policies. Furthermore, they have introduced or interpreted indicators (wrongly) assuming that globalization affects equally all countries. Thus, this represents an excellent playground to understand how context shapes the meaning of indicators. For instance, the increase of international collaboration since the 1980s (Adams, 2012) is usually interpreted as a positive factor for increasing scientific impact (Persson, Glänzel, & Danell, 2004). Mobility has also increased and is usually perceived as benefitting research careers (Sugimoto et al., 2017; Zippel, 2017). However there are contradicting views on whether the national impact of mobility is positive or negative (Arrieta, Pammolli, & Petersen, 2017; Levin & Stephan, 1999; Meyer, 2001). Since the end of 20th century many governments introduced publication policies that push researchers to publish in international venues (namely, the journals indexed in Web of Science and Scopus databases) and English language as a means to improve their profile internationally (Jiménez-Contreras, de Moya Anegón, & López-Cózar, 2003; Van Raan, 1997; Vessuri, Guédon, & Cetto, 2014).

These policies tend to be supported with indicators which are interpreted in the same manner – i.e. assuming that the more internationalisation and the more mobility, the better. For instance, an increase in international collaboration is assumed to be positively related to citation impact (Persson et al., 2004) and is especially encouraged in countries with lower national scientific impact (Bote, Olmeda-Gómez, & Moya-Anegón, 2013). Mobility is also considered as positive at the individual and global levels (Sugimoto et al., 2017; Wagner & Jonkers, 2017). However, it is perceived differently in specific countries, e.g. in Spain or China it is seen as positive when scientists' return is ensured (Andújar, Cañibano, & Fernandez-Zubieta, 2015; Jonkers & Tijssen, 2008), while in Africa it is perceive negatively due to the high risk of brain drain (Bassioni, Adzaho, & Niyukuri, 2016). Finally, publishing in English language is perceived as essential to improve the visibility of research outputs (Buela-Casal & Zych, 2012). In an influential piece, Leeuwen et al. (2001) proved the major biases against non-English languages journals in the Journal Impact Factor (JIF). Citation rates to these journals are consistently lower than in English language journals due to the lack of coverage to non-English literature in Web of Science. To counteract such bias, some journals from non-English countries have ceased publishing in their original language and turned into English with the expectation that this would increase their visibility and hence, their citation rates (Robinson, 2016).

These three examples (international collaboration, mobility and English publishing in non-English speaking countries) will be discussed in this chapter to showcase how a de-contextualised use of scientometric indicators can work against the implementation of policies seeking to improve national research systems.

The chapter is structured as follows. First, we frame the challenges raised by globalization of research, policies for internationalisation implemented in different countries and how this is shaping national scientific workforces. Next, we discuss two examples of where a de-contextualised use of indicators may lead to misinterpretations. These are the use of international collaboration to achieve greater scientific impact and the use of evaluation based on Journal Impact Factors to internationalise national scientific literature. Followed by this, we will discuss how a narrow interpretation of a global phenomenon such as the globalisation of the scientific workforce can lead to defining partial indicators which may be ill-suited. We conclude with some final remarks.

**Globalization and research evaluation**

Research has always had a fundamental global component attached to its endeavour. However, it is usually assumed that the dawn of the 21st Century marks the beginning of a 'truly' global scientific system (Altbach, 2004; Nicolas Robinson-Garcia & Jiménez-Contreras, 2017). Here we provide some evidence pointing in this direction. First, the rise of world university rankings with the launch of the Shanghai Ranking in 2003 (Aguillo, Bar-Ilan, Levene, & Ortega, 2010) unleashed a global competition for talent and resources (Hazelkorn, 2011). Despite their many and serious flaws, rankings have transformed the perceived prestige of universities (Bastedo & Bowman, 2010; Moed, 2017a) and have directly influenced decision making at the institutional level (Robinson-Garcia, Torres-Salinas, Herrera-Viedma, & Docampo, 2019, p. 233). Second, the shift from international scientific networks formed quasi-exclusively by western countries, to more inclusive global scientific collaboration networks (Wagner, Park, & Leydesdorff, 2015); derived partly from the R&D growth in countries such as China (Quan, Mongeon, Sainte-Marie, Zhao, & Larivière, 2019) or Brazil (Leta & Chaimovich, 2002). These new global communities are characterized by a tight and small core of countries in which dissemination of knowledge is dependable on a reduced number of countries (Leydesdorff & Wagner, 2008), while allowing the inclusion of new players in the global network (Wagner & Leydesdorff, 2005).

The transformation of the higher education landscape has confronted traditional universities with a new scenario. They are asked to respond to local problems and national priorities, while competing in and forming part of global scientific networks and responding to their expectations (Nerad, 2010). This dual challenge has directly influenced the development of scientometrics. Three examples are provided. First, the increasing interest on societal impact and interdisciplinary research ('Mode 2', Gibbons et al., 1994) has led to different proposals for measuring societal impact, in particular with 'altmetrics' (Díaz-Faes, Bowman, & Costas, 2019; Haustein, Bowman, & Costas, 2016; Robinson-Garcia, van Leeuwen, & Rafols, 2018), and developing indicators for measuring interdisciplinarity in research (Abramo, D'Angelo, & Costa, 2017; Larivière & Gingras, 2010; Leydesdorff & Rafols, 2011; Rafols, Leydesdorff, O'Hare, Nightingale, & Stirling, 2012). Second, the introduction of new public management methods in research management has led many governments and institutions to use indicators to assess individuals' careers within performance-based assessment systems (Ràfols, Molas-Gallart, Chavarro, & Robinson-Garcia, 2016), inducing a plethora of studies on individual research assessment (i.e., Costas, van Leeuwen, & Bordons, 2010; Hirsch, 2019). Third, the formation of international networks as a result of proactive policies has raised interest on the study of international collaboration (Bote et al., 2013; Leydesdorff & Wagner, 2008), and more recently, international mobility (Moed, Aisati, & Plume, 2013; Moed & Halevi, 2014; Robinson-Garcia et al., 2019; Sugimoto et al., 2017).

Scientometric indicators have grown in importance, in particular within national strategies of internationalisation. We will now briefly review some examples related with collaboration and publication venue. The globalization of science is often studied through the analysis of international co-authorship patterns and the structure of the networks that emerges from these patterns (Wagner, 2019). Despite some reservations raised (Persson et al., 2004), international collaboration is generally perceived as a positive factor to achieve a higher scientific impact and promote networks of prestigious scientists which may lead to more novel science (Wagner, 2019). As a consequence, it is common to observe its presence in world university rankings as well as to introduce mobility policies requiring scientists to return to the country of origin, so that they bridge between the receiving and sending countries (Fang, Lamers, & Costas, 2019).

Policies related with promoting certain publication strategies are well-known. They favour publishing in journals indexed in the Web of Science or Scopus, preferably in journals within the top quartile of Clarivate's Journal Citation Reports according to their Impact Factor. Countries implementing these types of policies presently or in the past include China (Quan, Chen, & Shu, 2017), Finland (Adam, 2002), Spain (Jiménez-Contreras, López-Cózar, Ruiz-Pérez, & Fernández, 2002), Czech Republic (Good, Vermeulen, Tiefenthaler, & Arnold, 2015) and major Latin American countries such as Mexico among others (Vessuri et al., 2014). As the Impact Factor is biased against non-English language (Leeuwen et al., 2001), scientists and national journals in non-English countries are pushed into publishing in English language (González-Alcaide, Valderrama-Zurián, & Aleixandre-Benavent, 2012) as a means of fostering internationalisation and with the expectation of gaining greater citation impact.

In all these cases, the rationale for introducing such policies is the same. International collaboration and publishing internationally (English language publications) is a signature of research quality that leads to high visibility. Science produced in this context (either through collaboration or by publishing in highly visible venues) leads to highly cited science. Finally, it is assumed that a system which produces more highly cited science is better (sometimes it may even be argued that it more positive and beneficial for society, e.g., Baldridge, Floyd, & Markóczy, 2004). But, as we will now discuss, context will shape to what extent this is true and the potential pitfalls of this type of argument. In the following section, we further explore the cases of international collaboration and publishing in English language. We will discuss some common examples on how a universal interpretation of scientometric indicators can be misleading depending on the context in which it is used.

**Two cases on how de-contextualized indicators lead to wrong interpretations**
*International collaboration*
International collaboration, measured by the share of publications in which affiliations from more than one country appear, is generally perceived as an intrinsically positive indicator. Thus, university rankings such as the World University Rankings, the Scimago Institutions Rankings or the Leiden Ranking, include the share of internationally co-authored publications as one of their dimensions. This perception is especially noticeable when discussing strategies for enhancing scientific development in countries outside the scientific core. For instance, Quan et al. (2019) state that:

*"For developing countries, international collaborations represent an ideal opportunity to improve both scientific visibility and research impact by allowing their researchers to work with colleagues from more advanced scientific countries"* (p. 708)

While this may be true to some extent regarding citation impact, it is questionable, at least when analysing a country's capacity to develop scientific knowledge autonomously and independently. In figure 1 we show the share of internationally co-authored publications for countries worldwide according to their income level. As observed, in the cases of high income, upper middle and lower middle countries, the shares are always below 40% of their total output. However, for low income countries, this share increases up to 86% of their total output, evidencing the fragility of their research systems and their dependence on developed countries when producing research outputs. Evidence shows that lower and middle income countries have a much higher citation impact when collaborating internationally. But does this mean that a 95% international collaboration rate is better than a 75% collaboration? At which point should policy foster the development of domestic capabilities without reliance on international collaboration? This questions to what extent can the same bibliometric indicators be either interpreted or even applied in developed and developing countries (Confraria, Mira Godinho, & Wang, 2017).

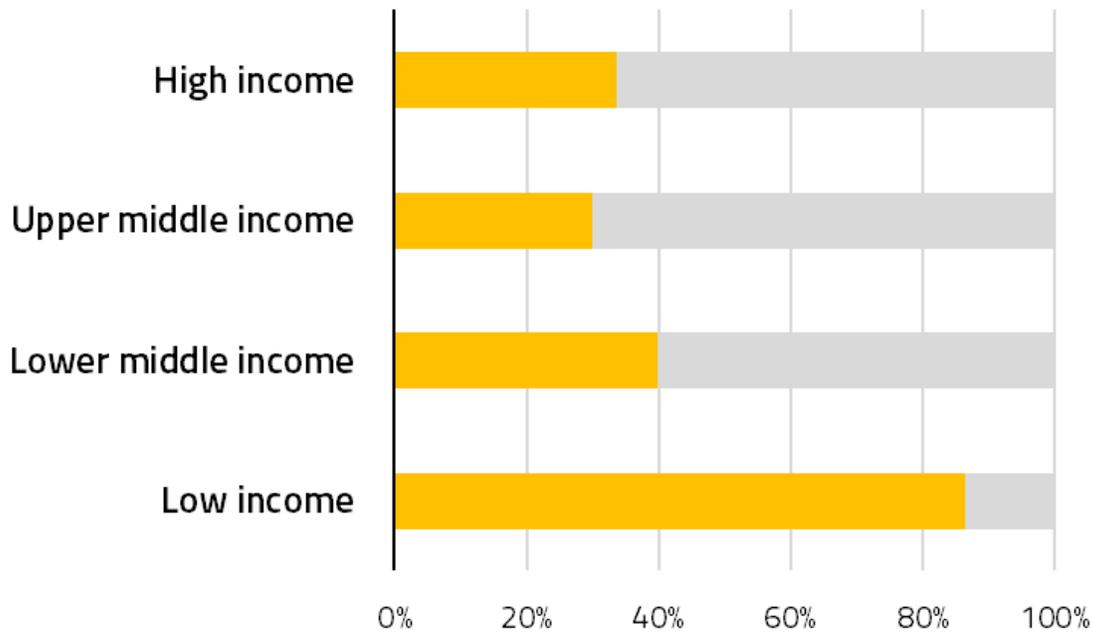

**Figure 1.** Share of internationally co-authored (yellow) and domestic (grey) publications indexed as articles in the Web of Science SCI, SSCI and A&HCI according to income country level (World Banks definition) in 2008-2017.

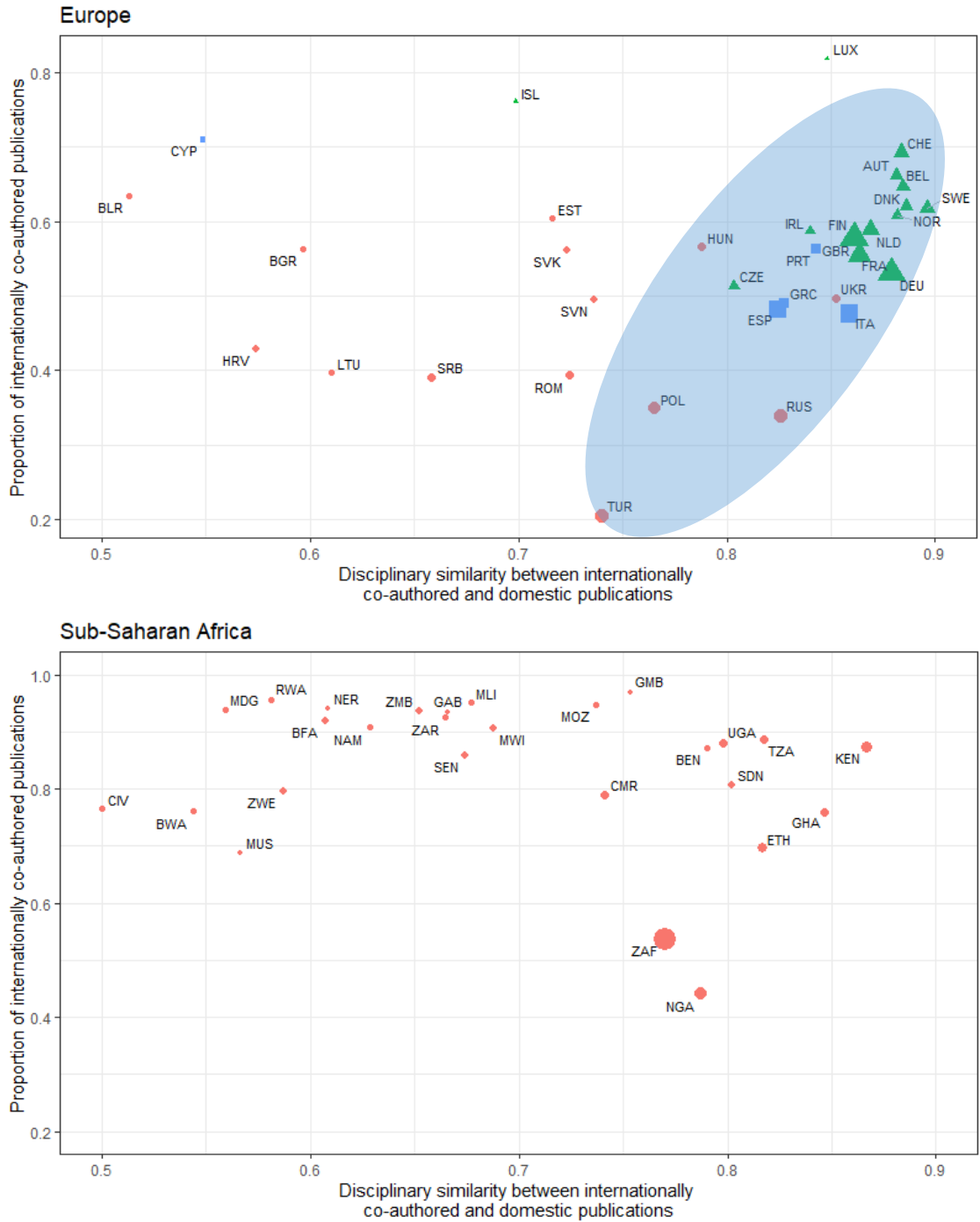

**Figure 2.** Scatterplots for countries from Europe & Central Asia (top) and Sub-Saharan Africa (bottom). X axis shows the proportion of internationally co-authored publications, y axis shows the cosine similarity of their disciplinary profile between their domestic and internationally co-authored publications. The size of a point reflects the total number of publications. Time period 1980-2018. Data from SCI, SSCI and H&CI. For European countries' colour and shape refers to region (red eastern European countries, blue southern European countries and green north and central European countries). Only countries with at least 8,000 publications are shown. For Sub-Saharan Africa only countries with at least

1,000 publications. Further information on the methodology is available in Robinson-Garcia, Woolley, et al. (2019)

In this context, it also seems reasonable to question if international collaboration can always best respond to local needs or it is driven by the pursue to integrate global scientific networks. In a case study focusing on six South Asian countries (Woolley, Robinson-Garcia, & Costas, 2017), we identified differences in the choice of partner and fields of interest when decoupling international collaboration between bilateral (co-authors affiliated to two distinct countries) and multilateral (co-authors affiliated to more than two distinct countries). We concluded that these two different collaboration patterns may be related to the nature of the fields as well as to the existence of mobility programs in which the emigrant bridges between countries while establishing wider networks with other countries. In a follow up study (Robinson-Garcia, Woolley, & Costas, 2019), we studied the degree to which countries follow global publication patterns using cosine similarity of the disciplinary profiles of countries with and without international collaboration. By combining these indicators, we can show that the interpretation of the indicator of proportion of international collaborations is not as straightforward as it could seem. Furthermore, context on the specific countries or regions is needed to better understand what is motivating such collaboration and if it seems reasonable to fit in with national interests.

The reason for this is that international collaboration, as conceived in scientometrics, is usually interpreted as a reciprocal relationship in which all partners are equal. However, in fact developed countries have positions of relative power in collaboration networks (Leydesdorff & Wagner, 2008), what causes asymmetries in the scientific partnerships (Chinchilla-Rodríguez, Bu, Robinson-García, Costas, & Sugimoto, 2018; Feld & Kreimer, 2019).

Figure 2 shows the disciplinary similarity of domestic versus internationally co-authored publications and the share of internationally co-authored publications for European (top) and African (bottom) countries. The case of Europe is of interest, as many policies have been put into place to coordinate the scientific integration of the different EU state members (Ackers, 2005). In this regard, we observe how northern and western countries tend to cluster together on the upper right side of the graph, correlating their domestic and international disciplinary profiles with their collaborating patterns (blue are in top chart in Figure 2). From the perspective of the European Research Area (ERA), this would be a desired path to follow and it would be expected for the rest of countries included in the plot, to align to this pattern. However, one might question if this would be the most appropriate choice from a national point of view, especially for eastern European countries, which tend to show a more dissimilar disciplinary profile.

In the case of Sub-Saharan Africa, the reading is completely different, and what we observe is a research profile completely overridden by international partners, with the exceptions of South Africa and Nigeria. This questions to what extent such research is based on local capabilities and responds to local demands and challenges. The very high share of internationally co-authored publications may actually be an indicator of the weakness of their national scientific systems and their dependence on international partners.

These two examples illustrate how the same indicator can be interpreted in different ways within (eastern European vs. western and north European countries) and between regions (Europe vs. Sub-Saharan Africa).

*Publishing in English as a strategy for internationalisation*

In this second case study we will focus on the influence of English language as a strategy to internationalize research in non-English speaking countries. We will focus on the share of publications in English language in non-English speaking countries. This indicator is perceived by researchers as a proxy for internationalising their research outputs (Buela-Casal & Zych, 2012). Here, citation-based indicators generally, and more specifically the Journal Impact Factor, are the indicators motivating such strategies. Already in 2001, Leeuwen et al. (2001) noted systematic biases in the Web of Science in terms of coverage of non-English literature and of the citation impact of such literature. Indeed, as shown in figure 3, English language accounts for 96% of the publications indexed, with none of the other 46 languages included surpassing 1% of the database (German, the second most common language, represented 0.9% of the database in the 2000-2017 period). Furthermore, despite the small drop suffered between 2006 and 2010, due to the inclusion of non-English journals in the database, English language rapidly caught up and even increased its share, representing 97.6% of the database in 2017.

However, this over-representation of English literature is seldom seen as a shortcoming of the data source. Instead, in many countries, lack of inclusion of national journals in the WoS is seen as evidence of the lack of internationalisation of their research outputs, as they are less visible (not indexed in these large databases) and less cited. This affects specially the Social Sciences and Humanities fields, as they are more prone to rely on national languages and address a more diversified audience than in other fields (Nederhof, 2006; Sivertsen, 2016).

But we do see also such negative connotation in other fields in which translational research and contact with practitioners is essential (Rey-Rocha & Martín-Sempere, 2012), such as Clinical Medicine. Despite the poor coverage of non-English literature, we observe in figure 4 (top chart) that in the case of Spain, there's been a huge flip from Spanish or other languages to English even within publications indexed in Web of Science (from more than 40% in non-English languages in 1980 to less than 20% in 2017). Furthermore, despite the efforts of these databases to include more non-English speaking literature, the weight on internationalisation seems to rapidly overcome such efforts. Figure 4 (bottom chart) shows the proportion of outputs from Brazil in the Social Sciences since 1980 to 2017. We observe a rapid turn into English in the last part of the 1980s and then an important increase of Portuguese literature in the second half of the 200os, due to the addition of Brazilian journals in the database. However, this increase is rapidly overridden and by 2015 we have the same proportion of English literature as we had before the inclusion of national journals.

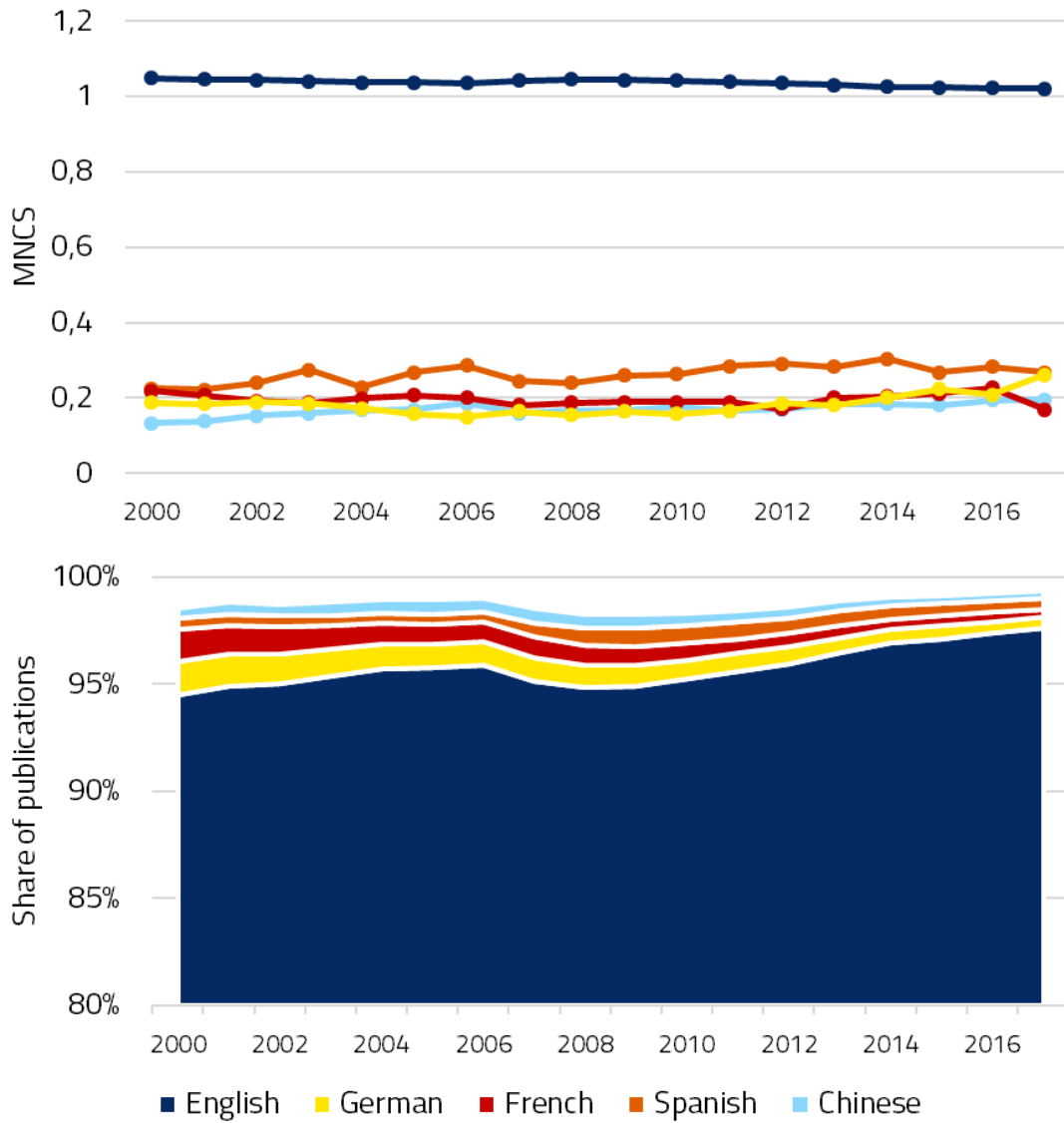

**Figure 3.** Mean Normalised Citation Score (MNCS at the top) and share of publications (bottom) of publications in the five most common languages available in the Web of Science for the 2000-2017 period.

This is due to the fact that many authors and journals from non-English speaking countries, switch their publications into English language, in the hope that they would achieve greater visibility and perhaps ensure higher citations. These strategies range from a complete switch to English, to maintaining bilingual versions of research articles or open in up to multilinguism, in which authors decide whether they wish to publish in their national language or English. Despite the overwhelming evidence on the citation advantage of English publications (see figure 3 and further evidences in González-Alcaide et al., 2012;

Liu, Hu, Tang, & Liu, 2018), experiences on turning into or adding English in a journal, have resulted in contradictory results (Robinson, 2016).

Larivière (2018) offers some possible reasons explaining why national journals may not succeed on increasing their impact when changing their publication language, or even if they manage to increase their impact, they never get similar Impact Factors to those achieved by journals from English speaking countries. First, there might be an author bias, as they tend to perceive national journals as less worthy and might decide to submit their lesser work to these journals. Second, these journals might focus on issues of local relevance which are not well covered by foreign journals (Piñeiro & Hicks, 2015). Also, these journals may have different functions than mainstream journals by serving as conduits to inform local communities (Chavarro, Tang, & Ràfols, 2017). For instance, recent correspondence in Nature raised awareness on the need to publish in non-English language to reach certain communities in India (Khan, 2019). Furthermore, forcing non-native speakers to publish in English presents a disadvantage with respect to native speakers, both for authors (Henshall, 2018) and journals (González-Alcaide et al., 2012), and may also lead to a duplication of research contents (both in their native and English language) to be able to reach national and international audiences, as observed in the case of Chinese literature (Liu et al., 2018).

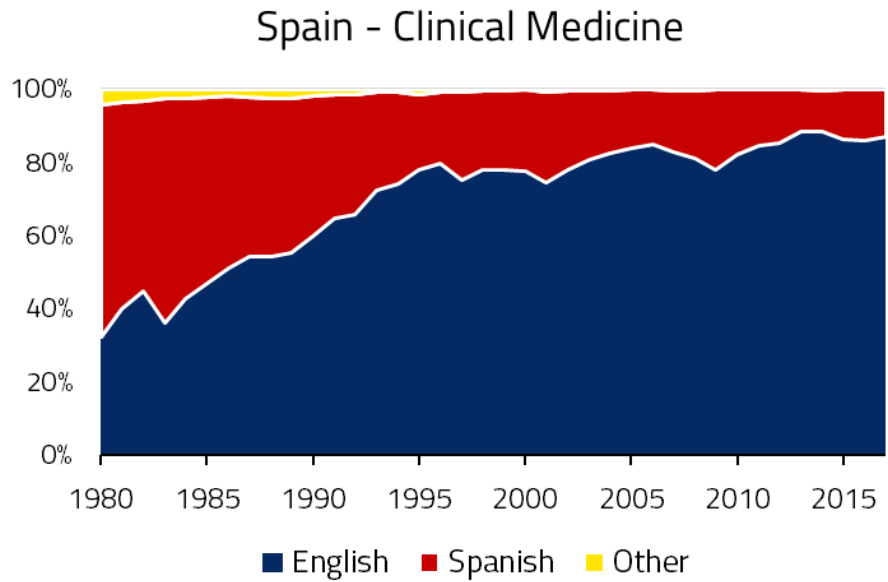

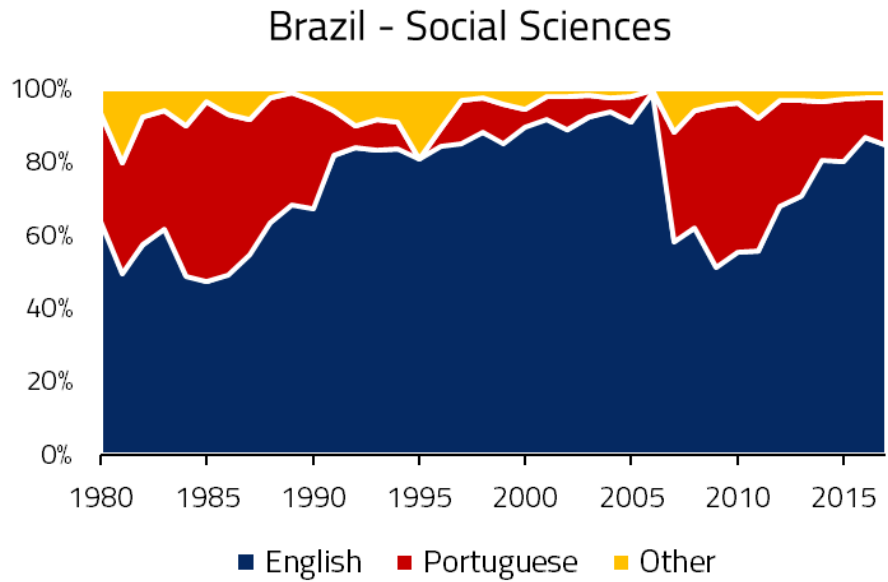

**Figure 4.** Proportion of publications in English, local and other languages for Spain in Clinical Medicine (top) and Brazil in Social Sciences (bottom) between 1980 and 2017.

Given such evidence, how internationalise one country's outputs without affecting their national visibility? Coming back to the dual challenge to which institutions are confronted in a global world, how can they open up their research findings to their local communities while becoming part of global scientific discussions? Sivertsen (2018) argues in this sense in favour of the promotion of multilinguism in science. He illustrates such notion with the case of Social Sciences and Humanities, commonly considered of a more localised nature (Hicks, 2005), but which can actually 'be valued as an example of combining international excellence with local relevance in a multilingual approach to research communication' (Sivertsen, 2018, p. 3). Therefore, the message is to shift away from what Neylon (2019) refers as false dichotomy,

*'that is the setting of local priorities towards societal engagement and wider impacts is positioned as being in opposition to "objective" and "international" measures of "excellence"'* (p. 4).

**Looking at international mobility from different angles**
In this section we now change the focus. Instead of looking at the interpretation of an indicator in different contexts, we will examine how different policy contexts referring to the same phenomenon, can lead to the selection of different indicators, that is, international flows of the scientific workforce. Mobility flows of scholars is a case in which universal uses of indicators clearly enter in conflict with the context in which they are applied. As we will now show, discussions on brain drain/gain or brain circulation tend to reflect different points of view of the same phenomenon, in which all interpretations tend to be partial. As Nerad (2010, p. 2) puts it:

> *" [D]ue to globalization, institutions responsible for graduate education today must fulfil a dual mission: building a nation's infrastructure by preparing the next generation of professionals and scholars for the local and national economy, both inside and outside academia, and educating their domestic and international graduate students to participate in a global economy and an international scholarly community. This dual mission is often experienced as a tension, because universities in many ways operate under a sole national lens."*

Such tension is reflected on the literature studying mobility flows. Now we will show two contrasting views of such phenomenon. This case differs from the two previous ones in the sense that here a national or regional view of the same phenomenon is affecting the election of the indicator used. We will now show two examples in which different indicators are used depending on the local policy context. On the one hand, the case of the European Union, where the emphasis is placed on brain circulation and the promotion of knowledge transfer among member states (Ackers, 2005). On the other hand, the United States, where interest in mobility is related with brain gain, or in other words, the capacity of attraction of highly skilled scientists (Levin & Stephan, 1999).

*The European Union and the promotion of international mobility*
In the case of Europe, interest on mobility derives from the desire to promote a stronger and more cohesive European Research Area (ERA), by developing strong knowledge flows and a common labour market that can compete with the US. Here bibliometric indicators have played a key role on the development of the Framework Programmes with CWTS leading such movement (Delanghe, Sloan, & Muldur, 2010). The promotion of knowledge flows within region is seen as one of the key strategies to ensure a more direct path towards innovation (Tijssen & van Wijk, 1999). This interest on mobility has permeated in many countries, fostering scientists to undertake short-term mobility periods (Cañibano, Fox, & Otamendi, 2016) and converting international experience as a pre-requisite in programmes attracting talent (i.e., Cañibano, Otamendi, & Andujar, 2008; Torres-Salinas & Jiménez-Contreras, 2015).

The scientometric study of the effects of such policies is usually based in the analysis of increases in international collaboration (Andújar et al., 2015; Jonkers & Tijssen, 2008), with few attempts at bibliometrically tracking mobility of scholars (Laudel, 2003). But the development of author name disambiguation algorithms recently allowed to bibliometrically track changes in affiliation of scientists, with Moed authoring the first large-scale analyses (Henk F. Moed et al., 2013). This has allowed to explore geographical, disciplinary or sectoral mobility flows of scientists (Moed & Halevi, 2014). Recent developments have expanded the notion of mobility by characterizing mobility types and mobility flows (Robinson-Garcia et al., 2019; Sugimoto, Robinson-Garcia, et al., 2017), and have open the door to better comprehending the relation between knowledge flows and mobility (Aman, 2018; Franzoni, Scellato, & Stephan, 2018) or characterising mobile scholars (Halevi, Moed, & Bar-Ilan, 2016a, 2016b).

Still, while being certainly novel and surpassing many of the limitations of previous attempts to study mobility flows (Sugimoto, Robinson-Garcia, & Costas, 2016), this type of approach offers a narrow definition of what mobile is, as it relies heavily on publication data (only if an author publishes in the origin and destiny countries can we define her as mobile) and ignores a very common type of mobility in Europe as is short term temporary mobility. This latter fact affects especially younger scholars who may move temporarily to other countries while retaining their home affiliation as part of their training.

*The United States and the attraction of foreign-born scientists*
A completely different body of literature can be found in the United States with regard to the globalization of the scientific workforce. In this country, foreign-born scientists represent 24% of the faculty (Lin, Pearce, & Wang, 2009) and their proportion keeps increasing, overcoming even new hires of domestic racial/ethnic minority groups (Kim, Twombly, & Wolf-Wendel, 2012). As observed, here the interest lies on how to attract and integrate foreign scholars who arrive to the country, as they have become a key asset for their national science system (Levin & Stephan, 1999; Lin et al., 2009). International experience is measured as an inherent characteristic of the individual given by their visa status and not by their experience of working in several countries. Hence, US born scientists are considered domestic and no interest is shown to their capacity to integrate global networks. Studies in this regard make use of scientometric indicators to analyse performance of this workforce (Stephan & Levin, 2001), but also rely on survey data, analysing other aspects such as job satisfaction (Mamiseishvili, 2011), productivity (Kim et al., 2012; van Holm, Wu, & Welch, 2019) or capacity of engagement with non-academic sectors (Libaers, 2014).

In these two examples, we observe how flows of scholars are defined differently. In the case of Europe, a 'global scholar' is someone who has changed their affiliation between countries, while in the case of the United States, is someone born abroad. When comparing both ways of operationalising the same phenomenon, we observe many disparities (Robinson-Garcia, van Holm, Melkers, & Welch, 2018). As a means to reconcile these two partial views of the same phenomenon, Welch et al. (2018) propose a comprehensive framework in which mobility or visa status are seen as two of the many features which characterize the global experience of scientists. They highlight the importance of considering such experience as a multi-layered concept. Also, these diverging ways of studying the effects of globalisation in the scientific workforce, highlight the importance of balancing between the local and global aspects of the same phenomenon in order to define indicators which are truly meaningful in a given national or regional science system.

**Concluding remarks**
In this chapter we have explored how the meaning of scientometric indicators can vary depending on the policy context in which they are applied. We have focused on the topic of globalisation of science and the use of scientometric indicators that countries make to support or introduce internationalisation policies. The chapter is inspired on Henk F. Moed's oeuvre in three different ways. First and most importantly, his conceptualisation and emphasis on the importance of policy context as a necessary framing when considering the use of scientometrics for research assessment (Moed & Halevi, 2015). Second, his work along with other colleagues on denouncing biases of the Journal Impact Factor, and the dangers of using it especially in non-English speaking countries and social sciences fields (Leeuwen et al., 2001; Moed & van Leeuwen, 1996). Finally, his pioneering work on the use of author name disambiguation algorithms to track mobility of scholars at large-scales (Moed et al., 2013).

By presenting two examples on how the use of scientometric indicators without considering the policy context in which they are applied, we have explored the interpretative ambiguity mostly ignored of widely used indicators, such as the share of internationally co-authored publications or the JIF. The purpose is not to condemn their use, as we do believe that the information provided by scientometric indicators is useful to inform policy managers. But to warn on the need to contextualise and use them as informative devices that can support research assessment exercises rather than as assessment devices which can be applied automatically. The third example illustrates a different case of misinterpretation. Here we show two cases in which the same phenomenon is being analysed using different indicators. From this case, we learn how a narrow view of a phenomenon which ignores the global policy context, can lead to short-sighted indicators which may not be adequate.